\documentclass[12pt]{article}

\usepackage{epsf,a4wide}

\newcommand{\eqcm}{\: ,} 
\newcommand{\eqpt}{\: .}

\begin{document}

\begin{flushright}
DAPNIA-SPHN-98-52 \\ CPHT-S637-0898 \\ hep-ph/9808479
\end{flushright}

\begin{center}
\vskip 3.5\baselineskip
\textbf{\LARGE Exclusive electroproduction of vector mesons \\
\medskip and transversity distributions}
\vskip 2.5\baselineskip
Markus Diehl$^{1}$, Thierry Gousset$^{2}$ and Bernard Pire$^{3}$
\vskip \baselineskip
1. DAPNIA/SPhN, CEA/Saclay, 91191 Gif sur Yvette, France \\
2. SUBATECH\,\footnote{Unit\'e mixte 6457 de l'Universit\'e de Nantes,
de l'Ecole des Mines de Nantes et de l'IN2P3/CNRS}, B.P. 20722, 44307
Nantes, France \\ 3. CPhT\,\footnote{Unit\'e mixte 7644 du CNRS},
Ecole Polytechnique, 91128 Palaiseau, France \\
\vskip 3\baselineskip
\textbf{Abstract} \\[0.5\baselineskip]
\parbox{0.9\textwidth}{We show that the leading twist contribution to
exclusive electroproduction of transversely polarized vector mesons
vanishes at all orders in perturbation theory. Therefore one cannot
extract information on the quark transversity distribution of the
nucleon from this reaction. In turn one has that the produced vector
meson is purely longitudinal in the large-$Q^2$ limit, which provides
a new test of the dominance of leading twist at finite $Q^2$.}
\vskip 1.5\baselineskip
\end{center}

\section{Introduction}

Exclusive electroproduction of photons or mesons off the nucleon
provides a powerful way to extract off-diagonal (off-forward,
nonforward) parton distributions~\cite{OFPD,NFPD}, making use of the
factorization properties of these processes at large
$Q^2$~\cite{CFS,fact}. In the diagonal limit these quantities are
related to the usual spin averaged, helicity or transversity dependent
parton distributions. The experimental difficulty of accessing the
chiral-odd (transversity) quark distribution
\begin{equation} \label{transversity}
h_1(x) \; \bar{u}(p,s') \sigma^{+i} u(p,s)=
{p^+\over2\pi} \int\! dz^-\, e^{i x (p^+ z^-)}\,
\langle p,s'|\, \bar{\psi}(0)\, \sigma^{+i}\, \psi(z^-)\, |p,s\rangle
\end{equation}
in inclusive or semi-inclusive processes~\cite{trans} has triggered a
strong interest in the related off-diagonal quantities~\cite{CFS,Hood}
which involve the matrix element $\langle
p',s'|\,\bar{\psi}(0)\,\sigma^{+i}\,\psi(z^-)\,|p,s\rangle$, where the
proton momentum in the bra and the ket is not the same.

\begin{figure}
\begin{center}
\leavevmode \epsfxsize 0.5\textwidth
\epsfbox{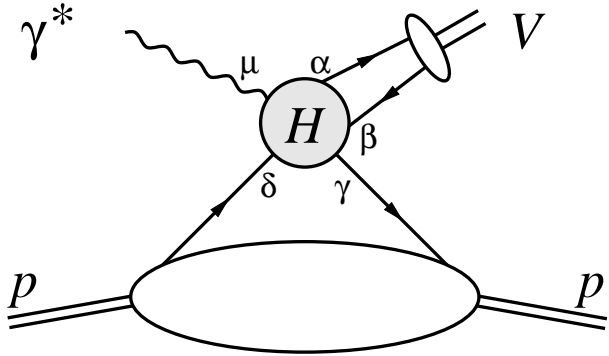}
\end{center}
\caption{\label{factorization} Factorization of $\gamma^\ast p\to Vp$
into an off-diagonal parton distribution, a hard photon-parton
scattering $H$, and the distribution amplitude of the produced meson.}
\end{figure}

In this paper we are concerned with vector meson production
$\gamma^\ast p \to V p$ in the kinematical regime where the photon
virtuality $Q^2$ and the squared c.m.\ energy $W^2$ are large while
the squared momentum transfer $t$ is small. In this Bjorken regime and
under the condition that the virtual photon has longitudinal
polarization the amplitude factorizes into an off-diagonal parton
distribution, a hard photon-parton scattering $H$, and the
distribution amplitude of the produced meson~\cite{CFS}, cf.\
Fig.~\ref{factorization}.

In reference~\cite{CFS} it has been noted that the production of a
transversely polarized vector meson involves the chiral-odd
off-diagonal parton distributions in the proton because the
distribution amplitude for a transversely polarized vector meson is
chiral-odd~\cite{BalBr,CZ} and because the trace of an odd number of
$\gamma$-matrices vanishes. It has however been observed~\cite{Mank}
that this contribution is zero at leading order in $\alpha_s$. In the
present short paper we confirm and generalize this result. That this
contribution vanishes is in fact due to angular momentum and chirality
conservation in the hard scattering and holds at leading power in
$1/Q$ to {\em all orders} in the strong coupling. In the large-$Q^2$
limit the produced meson must therefore have longitudinal
polarization; an analysis of its decay angular distribution thus
allows for stringent tests of the dominance of the leading
contribution in $1/Q$, which contains off-diagonal parton
distributions one would like to extract.

\section{The physical picture}

In the Bjorken regime we are interested in it is suitable to work in
the c.m.\ frame, where the initial and final state protons move fast
to the right and the produced meson fast to the left; we remark that
this kinematic situation is essential for the power counting arguments
in the factorization proof~\cite{CFS}.

Our argument has two main ingredients. The first is that the leading
term in $1/Q$ of the loop integral in Fig.~\ref{factorization} is
obtained by approximating the hard scattering $H$ by its value with
all partons being on shell and collinear along a suitably chosen
$z$-axis. For our purpose it is convenient to define this axis such
that the outgoing meson has zero transverse momentum, and to quantize
its spin along $z$. {}From rotation invariance the helicities of the
$q\bar{q}$-pair from the hard scattering must then add up to the
helicity of the meson. We remark in passing that on the proton side a
corresponding condition on the difference of initial and final proton
helicities only holds when the protons are strictly collinear.

The second ingredient is that the quark masses can be neglected in the
hard scattering so that the quark chirality is conserved in $H$; here
we make the usual assumption that in the limit of zero quark mass
there are no chirality breaking contributions to $H$ which would be
due to quantum effects.  Since the helicities of the quark and the
antiquark forming the transverse meson must be equal, the quark charge
flow in $H$ cannot connect the lines $\alpha$ and $\beta$ in
Fig.~\ref{factorization}. This excludes gluon exchange at the proton
side and we are left with a hard scattering $H$ with four external
quark legs and with the charge flow connecting line $\alpha$ with
$\delta$ and line $\beta$ with $\gamma$.

\newcounter{case}
\renewcommand{\theequation}{\arabic{equation}\alph{case}}

Depending on the parton momentum fractions, the hard scattering can
now be
\begin{eqnarray}
\gamma^\ast + q        &\to& q\bar{q} + q  \eqcm 
   \stepcounter{case} \label{q} \\ 
\gamma^\ast + \bar{q}  &\to& q\bar{q} + \bar{q} \eqcm 
   \stepcounter{case} \addtocounter{equation}{-1} \label{qbar} \\
\gamma^\ast + q\bar{q} &\to& q\bar{q} \eqcm 
   \stepcounter{case} \addtocounter{equation}{-1} \label{qqbar}
\end{eqnarray}
where we have omitted flavor indices. While cases (\ref{q}) and
(\ref{qbar}) are familiar from the usual diagonal parton
distributions, case (\ref{qqbar}) is characteristic of the kinematical
asymmetry between the two proton lines~\cite{NFPD,DG}.

\renewcommand{\theequation}{\arabic{equation}}

\begin{figure}
\begin{center}
\leavevmode \epsfxsize 0.5\textwidth
\epsfbox{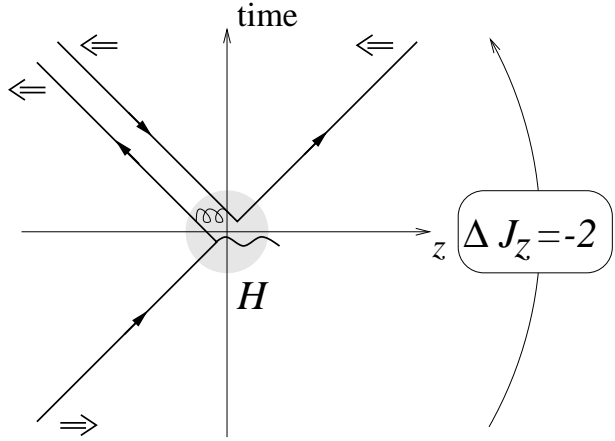}
\end{center}
\caption{\label{spacetime} Space time structure of the hard scattering
$\gamma^\ast q \to q\bar{q} + q$. The figure gives an example of the
hard scattering kernel at lowest order in $\alpha_s$ and of the
alignment of parton helicities along the $z$-axis.}
\end{figure}

In case (\ref{q}) the hard scattering has a space-time structure as
shown in Fig.~\ref{spacetime}. The right-moving quark from the proton
is scattered to become a left-moving quark in the meson, along with
the creation of a right-moving quark and a left-moving antiquark. For
a given helicity of the initial state quark all other parton
helicities are fixed from chirality conservation and the constraint to
form a transverse meson. Since the scattering is collinear the photon
would have to transfer two units of angular momentum as can be seen
from Fig.~\ref{spacetime}. This is of course not possible and we
conclude that the process cannot take place to leading order in
$1/Q$. In case (\ref{qbar}) the argument is completely analogous.

Along the same lines one can see that in case (\ref{qqbar}) the photon
would again have to transfer two units of angular momentum. We remark
that now the quark configuration in the hard scattering is the same as
in the reaction $\gamma^\ast V \to V$. {}From hadron helicity
conservation in the hard scattering mechanism~\cite{BL} it has long
been known that the elastic meson form factor for transverse
polarization is suppressed in powers of $1/Q$ compared with the
longitudinal one~\cite{CZ}.

\section{An algebraic proof}

The hard amplitude for the process is a tensor with one Lorentz index
and four Dirac indices (cf.\ Fig.~\ref{factorization}) and can be
decomposed as
\begin{equation}\label{first_decomposition}
H^{\mu}_{\alpha\beta\gamma\delta} = t^{\mu}_{mn \phantom{\delta\!\!}}\,
\Gamma^m_{\alpha\beta}\, \Gamma^n_{\gamma\delta} \eqcm
\end{equation}
where the $\Gamma^m \, (m=1,\dots,16)$ are the standard $4\times 4$
basis matrices in spinor space, $\Gamma^m=1$, $\gamma^5$,
$\gamma^{\lambda}$ ,$\gamma^5\gamma^{\lambda}$,
$\sigma^{\lambda\rho}$, and where a sum over $m$ and $n$ is
understood.

This expression is general but we want to focus on the leading term in
$1/Q$ of the amplitude (while retaining all orders in $\alpha_s$) for
which we have already noticed that the hard process is collinear along
the $z$-axis. Calling $t^{\mu}_{L\, mn}$ the leading term in a
$1/Q$-expansion of $t^{\mu}_{mn \phantom{\delta\!\!}}$ in
Eq.~(\ref{first_decomposition}) and introducing two lightlike vectors
$v=(1,0,0,1)/\sqrt{2}$ and $v'=(1,0,0,-1)/\sqrt{2}$, which
respectively define a plus and a minus direction, we thus have that
$t^{\mu}_{L\, mn}$ is a function of $v$ and $v'$,
\begin{equation}
t^{\mu}_{L\, mn} = f^{\mu}_{mn}(v,v') \eqcm
\end{equation}
but not of any transverse vector.

The next step is to impose the combination of quark helicities in
which we are interested. One way of finding which matrices $\Gamma^m$
this selects on the meson side in Eq.~(\ref{first_decomposition}) is
to contract $H$ with massless spinor solutions of the Dirac equation
whose momenta point in the minus direction and whose helicities are
aligned, i.e.\ both positive or both negative. Since the combination
of these spinors is
\begin{equation}
v_{\beta}\, \bar{u}_{\alpha} = a_i \, \sigma^{+i}_{\beta\alpha} \eqcm
\end{equation}
where $a_i$ is a vector in the transverse plane, we find that the
selected $\Gamma^m$ are $\sigma^{-i}$ with $i=1,2$. This corresponds
to the fact that the leading twist distribution amplitude for a
transversely polarized vector meson is constructed from the matrix
element
\begin{equation}
\langle V|\bar{\psi}(z^+)\, \sigma^{-i} \psi(0)|0\rangle \eqpt
\end{equation}
We remark that this Dirac matrix structure is due to a selection on
helicity and not particular to a vector meson final state. Thus the
leading twist $q\bar{q}$-amplitude also goes with $\sigma^{-i}$ if we
have, for instance, a left moving tensor meson with helicity $\pm
1$. In Sect.~2 we have seen that we must route the parton chiralities
from the meson to the proton side, and therefore we have to carry out
the same exercise for the proton, with the interchange of $+$
and~$-$. For $\Gamma^n$ this selects $\sigma^{+j}$, where $j=1,2$ is
again an index in the transverse direction. This is precisely the
Dirac matrix structure entering in the definition of the transversity
distribution in Eq.~(\ref{transversity}).

Among the $16\times 16$ combinations of $m$ and $n$ in $t^{\mu}_{L\,
mn}$ we thus only need the $2\times2$ combinations $t^{\mu}_{L\, ij}$
going with $\sigma^{-i}_{\alpha\beta}$ and
$\sigma^{+j}_{\gamma\delta}$. Lorentz invariance requires the tensor
$t^{\mu}_{L\, ij}$ to be constructed from $v^{\mu}$, $v'^{\mu}$ and
the transverse tensors $g_{T\, ij}^{\phantom{\mu}}$ and $\epsilon_{T\,
ij}^{\phantom{\mu}}$ defined as
\begin{equation}
g_T^{\lambda\rho}=g^{\lambda\rho}-v^{\lambda}v'^{\rho}
                 -v'^{\lambda}v^{\rho} \eqcm \hspace{3em}
\epsilon_T^{\lambda\rho}=\epsilon^{\lambda\rho\sigma\tau} 
                         v^{\phantom{'}}_{\sigma} v'_{\tau} \eqpt
\end{equation}
Parity invariance would further restrict $t^{\mu}_{L\, ij}$ to be
proportional to $g_{T\, ij}^{\phantom{\mu}}$, but we do not need this
for our argument.

Now we want to impose chirality conservation. To this end we perform a
Fierz transformation,
\begin{equation}\label{second_decomposition}
H^{\mu}_{\alpha\beta\gamma\delta} = \widetilde{t}\,^{\mu}_{mn}\,
\Gamma^m_{\alpha\delta}\,\Gamma^n_{\gamma\beta} \eqcm
\end{equation}
and notice that chirality conservation means that both fermion lines
$\alpha\delta$ and $\gamma\beta$ consist of an odd number of
$\gamma$-matrices.\footnote{This is readily seen diagrammatically:
attaching $n$ gluons (or $n-1$ gluons and a photon) to a quark line
one obtains $n$ $\gamma$-matrices from the vertices and $n-1$ from the
massless quark propagators in between.} In the decomposition
Eq.~(\ref{second_decomposition}) $\Gamma^m$ and $\Gamma^n$ are thus
either $\gamma^{\lambda}$ or $\gamma^5\gamma^{\lambda}$. To leading
power in $1/Q$ and for aligned helicities of the final $q\bar{q}$-pair
we then have
\begin{equation}
\label{traces}
\widetilde{t}\,^{\mu}_{L\, mn} =
t^{\mu}_{L\, ij} \; \mathrm{tr}
\Big(\Gamma_m \sigma^{-i}\Gamma_n\sigma^{+j}\Big) \eqcm
\end{equation}
with $\Gamma_m$ and $\Gamma_n$ restricted to $\gamma^{\lambda}$ and
$\gamma^5 \gamma^{\lambda}$. Here $\Gamma_m$ with a lower index is
defined by $\mathrm{tr} (\Gamma^{m} \Gamma_{n})= \delta^{m}{}_{\!n}$,
where $\delta^{m}{}_{\!n}$ is the Kronecker symbol. Up to factors the
traces in (\ref{traces}) are given by
\begin{eqnarray}  \label{long-traces}
\mathrm{tr} \Big(\gamma^\lambda \sigma^{-i}
                 \gamma^\rho \sigma^{+j}\Big) \phantom{\gamma^5}
&\propto& g_T^{\lambda\rho \phantom{j\!\!}} g_T^{ij} - 
          g_T^{\lambda i \phantom{j\!\!}} g_T^{\rho j} - 
          g_T^{\lambda j \phantom{j\!\!}} g_T^{\rho i} 
\eqcm \nonumber \\
\mathrm{tr} \Big(\gamma^\lambda \sigma^{-i}
                 \gamma^\rho \sigma^{+j} \gamma^5\Big)
&\propto& g_T^{\lambda i \phantom{j\!\!}} \, \epsilon_T^{\rho j} + 
          \epsilon_T^{\lambda i \phantom{j\!\!}} \, g_T^{\rho j} \eqcm
\end{eqnarray}
where we have used that $i,j=1,2$. They vanish when contracted with
$t^{\mu}_{L\, ij}$ so that all in all we have
\begin{equation} \label{zero}
\widetilde{t}\,^{\mu}_{L\, mn} = 0
\end{equation}
for $m,n=1,\ldots,16$.

We remark that the expressions (\ref{long-traces}) are nonzero for any
fixed pair $i$, $j$, and that zero is only obtained by the contraction
with $g_{T\, ij}$ or $\epsilon_{T\, ij}$. To get to (\ref{zero}) we
are thus using the invariance of the hard scattering under rotations
around the $z$-axis; this corresponds to angular momentum conservation
along $z$ which went into our argument in Sect.~2.

\section{Phenomenology}

Let us first stress that the above derivation is valid for any spin
state of the incoming virtual photon. Phenomenological tests for the
suppression of transverse meson polarization are thus different from
and complementary to those for the dominance of the longitudinal
component of the virtual photon, which is part of the factorization
theorem~\cite{CFS}. The photon polarization can be tested using the
distribution in the angle between the hadronic and leptonic planes in
the electroproduction process $e p \to e V p$, which to leading order
in $1/Q$ should be flat~\cite{DGPR}.

The fact that the production amplitude for a transverse vector meson
vanishes to leading power in $1/Q$ translates into the
$Q^2$-dependence of its helicity density matrix elements
$\rho_{\lambda\lambda'}$ and thus of its decay angular
distribution. The general expression~\cite{mesons} for the decays
$\rho \to \pi \pi$, $\phi \to K K$,
\begin{eqnarray}
{1\over N} {dN \over d\!\cos\theta \, d\phi} &=& {3\over4\pi}
   \Big[ {\textstyle\frac{1}{2}} (1-\rho_{00})
   + {\textstyle\frac{1}{2}} \cos^2\theta \, (3\rho_{00} -1)
   \nonumber \\
&& {}- {\textstyle\frac{1}{\sqrt{2}}} \sin 2\theta \cos\phi \, 
       (\mathrm{Re}\,\rho_{10} - \mathrm{Re}\,\rho_{-10})
     + {\textstyle\frac{1}{\sqrt{2}}} \sin 2\theta \sin\phi \, 
       (\mathrm{Im}\,\rho_{10} + \mathrm{Im}\,\rho_{-10})
   \nonumber \\
&& {}- \sin^2 \theta \cos 2\phi \, \mathrm{Re}\,\rho_{1-1}
     + \sin^2 \theta \sin 2\phi \, \mathrm{Im}\,\rho_{1-1} \Big]
   \eqcm
\end{eqnarray}
will thus become
\begin{equation}
{1\over N} {dN \over d\!\cos\theta \, d\phi} = 
{3\over4\pi} \, \cos^2\theta + O\left(1\over Q\right)  \eqpt
\end{equation}
in the scaling regime.

As remarked in Sect.~3 our derivation also holds for a transversely
polarized meson with higher spin. The $q\bar{q}$-pair produced in the
hard scattering cannot have total helicity $\pm 1$ at leading order in
$1/Q$ and thus must have helicity zero. In the scaling limit the meson
is therefore again produced with longitudinal polarization and its
decay angular distribution completely determined.

Going a step further we can also apply our arguments to the production
of a nonresonant state of two spinless mesons such as $\gamma^\ast p
\rightarrow \pi \pi p$ or $\gamma^\ast p \rightarrow K K p$ when the
invariant mass of the two mesons is much smaller than the photon
virtuality, provided one replaces the meson distribution amplitude by
the generalized distribution amplitude introduced
in~\cite{DGPT}. While the nonresonant meson pair is no longer produced
in a single partial wave its angular momentum along the $z$-axis is
still zero in the scaling limit: the distribution in the polar angle
$\theta$ is then no longer given generically, but the dependence on
the azimuth $\phi$ is flat. This test is the precise analog of the one
for the dominant photon helicity mentioned above.

There is an important program of studying exclusive meson production
in ongoing or planned experiments~\cite{experiments,plans}. We stress
that the polarization structure of this reaction in the Bjorken regime
offers a valuable tool to test the dominance of the leading mechanism
in $1/Q$, which is essential before leading-twist quantities such as
off-diagonal parton distributions can be extracted.

\section*{Acknowledgments} We gratefully acknowledge discussions
with J.P. Ralston and O.V. Teryaev. This work has been partially
funded through the European TMR Contract No.~FMRX-CT96-0008:
Hadronic Physics with High Energy Electromagnetic Probes.

\end{document}